\documentclass[english,a4paper,pre,superscriptaddress,twocolumn,amsmath,amssymb]{revtex4}
\usepackage{graphicx,color,soul}

\newcommand{\mean}[1]{\langle #1\rangle}
\newcommand{\unit}[1]{\;\mathrm{#1}}
\newcommand{\mat}[1]{\boldsymbol{#1}}

\newcommand{\di}{\text{d}}
\newcommand{\ree}{R_\text{ee}}
\newcommand{\rpl}{R_\text{pull}}

\def\cal{calix[4]arene }

\begin{document}

\author{Fabian Knoch}
\affiliation{Institut f\"ur Physik, Johannes Gutenberg Universit\"at Mainz, Staudingerweg 7-9, 55128 Mainz, Germany}
\author{Ken Sch\"afer}
\affiliation{Institut f\"ur Physikalische Chemie, Johannes Gutenberg Universit\"at Mainz, Duesbergweg 10-14, 55128 Mainz, Germany}
\author{Gregor Diezemann}
\affiliation{Institut f\"ur Physikalische Chemie, Johannes Gutenberg Universit\"at Mainz, Duesbergweg 10-14, 55128 Mainz, Germany}
\author{Thomas Speck}
\affiliation{Institut f\"ur Physik, Johannes Gutenberg Universit\"at Mainz, Staudingerweg 7-9, 55128 Mainz, Germany}

\title{Dynamic coarse-graining fills the gap between atomistic simulations and experimental investigations of mechanical unfolding}

\begin{abstract}
  We present a dynamic coarse-graining technique that allows to simulate the mechanical unfolding of biomolecules or molecular complexes on experimentally relevant time scales. It is based on Markov state models (MSM), which we construct from molecular dynamics simulations using the pulling coordinate as an order parameter. We obtain a sequence of MSMs as a function of the discretized pulling coordinate, and the pulling process is modeled by switching among the MSMs according to the protocol applied to unfold the complex. This way we cover seven orders of magnitude in pulling speed. In the region of rapid pulling we additionally perform steered molecular dynamics simulations and find excellent agreement between the results of the fully atomistic and the dynamically coarse-grained simulations. Our technique allows the determination of the rates of mechanical unfolding in a dynamical range from approximately $10^{-8}$/ns to $1$/ns thus reaching experimentally accessible time regimes without abandoning atomistic resolution.
\end{abstract}

\maketitle


\section{Introduction}

The unfolding pathway of biomolecules and molecular complexes can be studied on a single molecule level by applying mechanical forces giving detailed information about conformational transitions in soft matter 
systems~\cite{Evans:2001,Zoldak:2013,Dudko:2015}.
The most prominent methodology to probe the response to a force is to keep one end of the macromolecule fixed and to pull on the other end with a constant velocity.
This way, the characteristic forces required to induce conformational changes can be recorded with high precision and information about the energy landscape and the relevant transition states can be gathered~\cite{Hyeon:2007}.
Using this information allows to test theoretical concepts like nonequilibrium fluctuation theorems and the laws of stochastic thermodynamics~\cite{Schmiedl:2007,Jarzynski:2011,Gupta:2011,Bustamante:2005,Alemany:2010}.
Additionally, using stochastic models of diffusive barrier crossing, it is possible to obtain relevant kinetic information by transforming measured rupture forces into kinetic rates~\cite{Dudko:2008,Maitra:2010,Zhang:2013,Bullerjahn:2014,Nam:2015}.
As in many areas of chemical and biological physics, in addition to experimental investigations computer simulations provide detailed information about relevant structural arrangements, statistical mechanical properties and kinetics on an atomistic level~\cite{Freddolino:2010,Luitz:2015}.
However, atomistic simulations of the unfolding of all but the smallest biomolecules still are challenging because of the rather long time scales involved and the system size to be considered~\cite{Hashem:2009,Perilla:2015}.
To overcome the resulting limitations a number of coarse-graining methodologies have been developed that allow to study important biophysical processes on experimentally relevant time scales~\cite{Saunders:2013,Fogarty:2015}.
While most of the coarse-grained models employ simplified interaction potentials and thus reduce the computational cost, the strategy of Markov state models (MSMs) consists in using the dynamical information from short atomistic runs to extrapolate to long time scales~\cite{Prinz:2011,Schwantes:2014,Hummer:2015,Wu:2016}.

The mechanical unfolding of molecules and assemblies can be studied computationally via steered molecular dynamics (SMD) 
simulations~\cite{Grubmuller:1996,Isralewitz:2001,Chen:2011,de2016}.
Such simulations allow to study unfolding pathways in atomistic resolution, inaccessible in experiments.
Additionally, they can be used to compare mechanical and chemical unfolding pathways~\cite{Stirnemann:2014} and also to compute the potential of mean force with the pulling direction serving as an order parameter~\cite{Park:2004}.
When compared to experimental force spectroscopy, the most prominent difference is given by the fact that experimental pulling velocities, typically in the range of $10^{-9}-10^{-5}$ m/s, are many orders of magnitude smaller than those used in SMD simulations, which usually vary between between $10^{-2}$ m/s and $10^{3}$ m/s~\cite{Zoldak:2013,Lee:2009,Habibi:2016}.
Only in recent years has it become possible to increase the experimental pulling velocities to values on the order of $10^{-3}$ m/s, thus reaching the range of SMD simulations~\cite{Lee:2009,Rico:2013}.
Still, a direct comparison between simulation results and experimental investigations is out of reach and in particular the conduction of a large number of SMD trajectories allowing a sound statistical analysis remains challenging.

\begin{figure}[t]
  \centering
  \includegraphics[width=.9\linewidth]{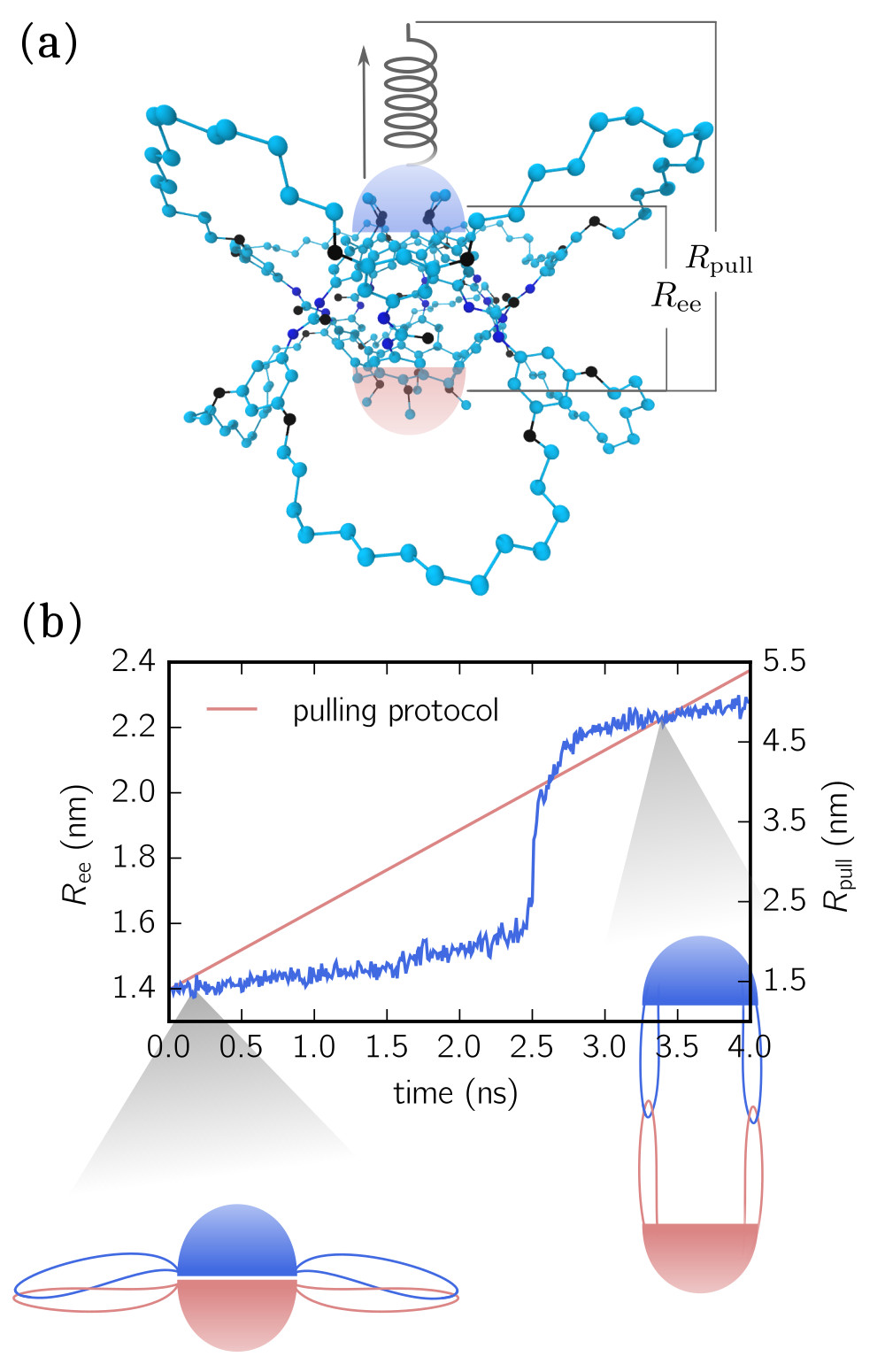}
  \caption{\textbf{Steered molecular dynamics simulations.} (a)~Atomic structure of calix[4]arene. The reference and pulling group are highlighted by a red and blue cup, respectively. During the simulation, the pulling group is separated from the reference group by a moving harmonic potential. (b)~End-to-end distance $\ree(t)$ and pulling protocol $\rpl(t)$ for $v=1\unit{m/s}$. The cartoon of the \cal dimer shows only two of the four aliphatic loops for clarity.}
  \label{fig:cal}
\end{figure}

Alternatively, the experimental conditions of force spectroscopy can be met routinely when performing simulations using coarse-grained models for the interactions and in particular if the coarse-graining procedure includes the use of implicit solvent models~\cite{Habibi:2016,Hyeon:2006,Best:2008}.
Such simulations yield important insights into the mechanical unfolding processes, however, they lack atomistic resolution.
Apart from the question of a pulling velocity dependence of the unfolding pathway~\cite{Li:2009}, one has to deal with the fact that the dynamics of coarse-grained models usually is faster than the atomistic one~\cite{Depa:2011}.
In the present work, we therefore develop a different methodology by using the information we obtain from detailed computations of the potential of mean force using the Umbrella method as a function of a pulling coordinate~\cite{Torrie:1977,Roux:1995}.  We build MSMs to simulate the transitions among the different configurations and switch between configurations using a force ramp protocol with the pulling device stiffness given by the curvature of the Umbrella potential. As a test case for our strategy we use a well studied calix[4]arene catenane dimer that shows two state behavior with the transitions among a ``closed'' and a ``open'' configuration associated with the reorganization of a hydrogen bond network~\cite{G69,G74,G77}, see Figure~\ref{fig:cal}(a). The catenane structure with four interleaved aliphatic loops prevents the calixarene ``cups'' to separate completely and thus allows to study two-state behavior in detail.

\section{Methods}

\subsection{MD simulations}
\label{sec:sim}

All molecular dynamics simulations are performed employing the Gromacs 5.1.2 software package~\cite{gromacs5}. The calix[4]arene dimer is positioned in a $5.7\times5.7\times 5.7~$nm box with periodic boundary conditions and filled with 827 mesitylene molecules. Molecular interactions are described by the OPLS force field~\cite{jorgensen1996}. All simulations are conducted in the NPT ensemble, for which the temperature is set to 300~K employing the velocity-rescaling thermostat~\cite{bussi2007}. To maintain a constant pressure of one bar, the Parrinello-Rahman~\cite{parrinello1981} barostat is used with time constant $2~$ps and compressibility of $8.26\times10^{-5}~\text{bar}^{-1}$. Long-range electrostatics are treated using particle mesh Ewald summation method~\cite{ewaldsum}, while for van der Waals interactions a dispersion correction~\cite{allen1989} is applied. The cutoff for all short-ranged interactions is set to 1.4~nm. All hydrogen involving covalent bonds are constraint by the LINCS algorithm~\cite{lincs}, allowing a time step of 2~fs.

\subsection{Markov state modeling}

\begin{figure}[b!]
  \centering
  \includegraphics[width=\linewidth]{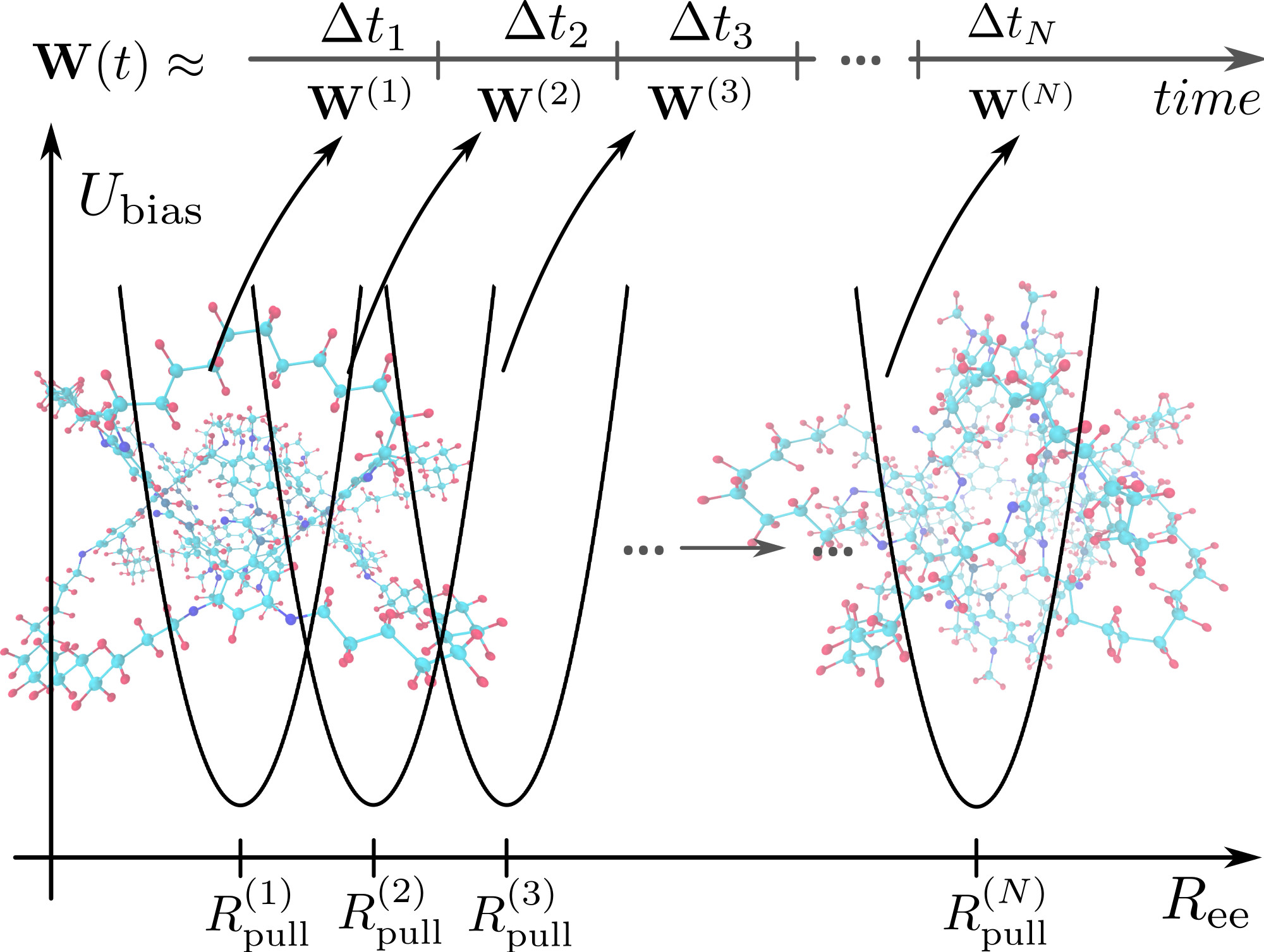}
  \caption{Sketch illustrating the model building procedure. For each simulated (constant) biasing potentials the corresponding time-independent rate matrix $\mat W^{(k)}$ is estimated. The time-dependent rate matrix $\mat W(t)$ (see eq.~\eqref{eq:master_approx}) is then constructed from these estimations.}
  \label{fig:msm}
\end{figure}

We discretize the configuration space into $N=60$ states with coordinates $\ree^i$ positioned equally spaced between $1.35~$nm and $2.2~$nm. With these states, we build 41 MSMs, each at another value of $2.0\unit{nm}\le\rpl\le4.0\unit{nm}$ apart 0.05~nm. For each biasing potential $U^{(k)}$ we conduct 40-100 MD simulations with an individual length of 20~ns. All simulations are randomly initialized from SMD ($v = 0.01\unit{m/s}$), while we discard the first 2~ns of each trajectory. The stiffness of the biasing potential is set to $\kappa=500~$kJ/(mol$\cdot\text{nm}^2$). To obtain the transition probabilities $\mat T^{(k)}=(T^{(k)}_{ij})$ of each MSM, we employ the TRAM estimator~\cite{wu2016} with lag time 1~ns according to a lag time analysis. This estimator combines the data from all simulations for all potentials to improve the estimation of the transition probabilities, in particular for rare transitions. From $\mat T^{(k)}$ one can infer the steady-state probability distribution $\mathbf p^\text{ss}$ (and thus the free energy landscape $F_i=-k_\text{B}T\ln p^\text{ss}_i$) by computing the eigenvector that is associated with the largest eigenvalue of unity. In this study we omit error estimates for the Markov state modeling approach since no rigorous estimation technique for the TRAM estimator exists yet.

\subsection{Time-dependent master equation}

Due to the force ramp changing $\rpl$, we are concerned with a time-dependent process obeying the continuous time master equation
\begin{equation}
  \label{eq:master}
  \partial_t \mathbf p^T(t) = \mathbf p^T(t)\mat W(t).
\end{equation}
Here the row-stochastic rate matrix $\mat W(t)$ (with $W_{ij} \ge 0 $ for $i\neq j$ and $\sum_{j} W_{ij} =0$) appears. We approximate this matrix by the sequence of rate matrices $\mat W^{(k)}$ corresponding to each MSM. These are obtain from $\mat T^{(k)}$ employing the series expansion of the matrix logarithm~\cite{israel2001}
\begin{multline}
  \mat W^{(k)} = \frac{1}{\tau}\log\mat T^{(k)} = \frac{1}{\tau}\log{(\mat 1 + \mat T^{(k)} - \mat 1)} \\ \approx \frac{1}{\tau}\left[(\mat T^{(k)} - \mat 1) - \frac{(\mat T^{(k)} -\mat 1 )^2}{2} + \frac{(\mat T^{(k)}-\mat 1)^3}{3} - ... \right].
\end{multline}
Here $\mat 1$ is the unity matrix and $\tau=1~$ns is the lag time for which $\mat T^{(k)}$ has been estimated. The series expansion is truncated if the next expansion term is small enough or the rate matrix $\mat W^{(k)}$ turns nonphysical, \emph{i.e.}, any nondiagonal element of $\mat W^{(k)}$ becomes negative. The formal integration of Eq.~\eqref{eq:master} leads to
\begin{equation}
  \label{eq:master_approx}
  p_i(t) \approx p_i(0) + \sum_{k=0}^{K-1} \int_{t_k}^{t_{k+1}} \di t' \sum_j 
  p_j(t')W^{(k)}_{ji},
\end{equation}
which we solve using a standard ODE solver.

\section{Results}

For the calix[4]arene dimer, we define two groups of atoms with $\ree$ denoting the fluctuating distance between the center-of-mass of each group [Figure~\ref{fig:cal}(a)]. To the ``pulling group'' a force is applied through the potential $U=\frac{\kappa}{2}(\ree-\rpl)^2$ with stiffness $\kappa$. In contrast to the ``force clamp'' mode pulling with a constant force, this potential holds the distance $\rpl$ constant while $\ree$ fluctuates.
The position of this potential is then changed with constant speed $v$ so that $\rpl(t)=\rpl^{(0)}+vt$, which is also called ``velocity clamp'' mode.

For large pulling speeds we perform SMD simulations. An exemplary trajectory illustrating the pulling protocol for $v=1~$m/s is shown in Figure~\ref{fig:cal}(b). In the beginning, the end-to-end distance $\ree$ follows the pulling protocol approximately linearly, although with a smaller slope. At $t\approx2.5$~ns, $\ree$ jumps to a larger value, indicating that the cuplike structures of the calix[4]arene dimer are largely separated. After this sharp transition, $\ree$ grows again linearly with the pulling protocol. The calix[4]arene dimer is referred to as being in the closed state when both cuplike structures are close together, and being in the opened state if they are largely separated. The abrupt jump in $\ree$ marks the transition between both states. In the following, we define $\tau^\ast$ as the transition time for the open/closed transition of the forced catenane dimer. Schlesier~\emph{et al.}~\cite{G77,schlesier2011} found the sharp transition of the end-to-end distance to be consistent with the rupture and formation of hydrogen bonds between urea groups and between urea and ether oxygen groups, respectively. Moreover, via conducting a large number of independent simulations they showed that the transition is indeed a stochastic event.

\begin{figure*}[t]
  \centering
  \includegraphics[width=.8\linewidth]{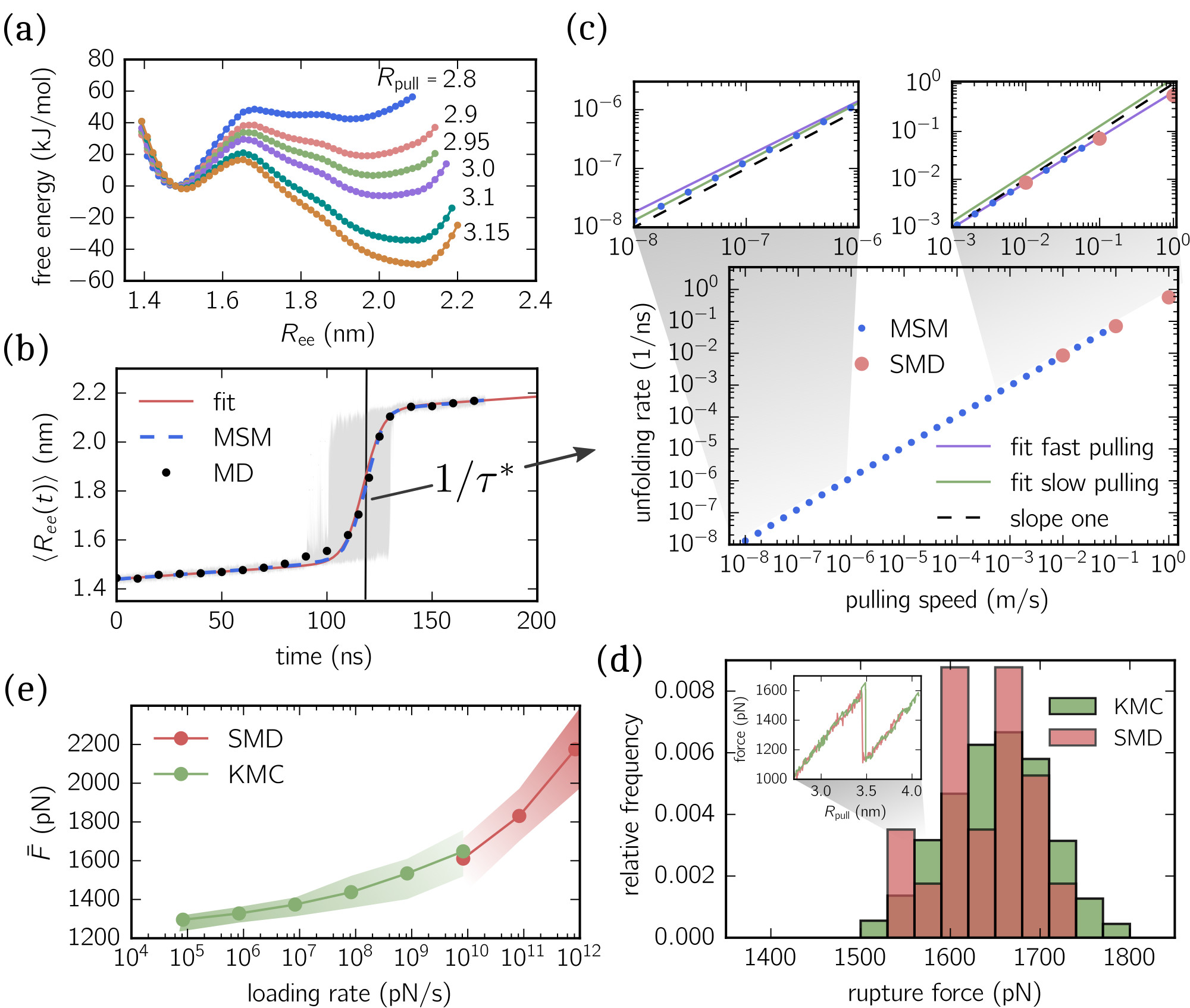}
   \caption{\textbf{Free energy landscape and unfolding dynamics.} (a)~Biased free energy profiles (potentials of mean force) as a function of end-to-end distance for different values of $\rpl$ (in nm). (b)~Mean end-to-end distance $\mean{\ree(t)}$ as a function of time calculated from 22 SMD simulation runs (black dots) and from our Markov state modeling approach (blue line) for pulling speed $v = 0.01~$m/s. The gray area indicates the $\sigma$-confidence interval computed from the SMD runs at each time. We extract the average unfolding time $\tau^\ast$ (vertical line) by fitting the data (red line) with the nonlinear expression given in Eq.~\eqref{eq:fit}. (c)~Unfolding rate $1/\tau^\ast$ for different pulling speeds. Upper panels show the detailed unfolding rates for slow (left panel) and fast (right panel) pulling speeds. Solid lines represent fits for fast ($v>10^{-3}\unit{m/s}$, purple) and slow ($v<10^{-6}\unit{m/s}$, green) speeds, while the black dotted line has unity slope. (d)~Comparison of constant-speed rupture force distributions for $v=0.01~$m/s. The red distribution is calculated from 22 SMD simulations, whereas the green distribution is obtained from kinetic Monte Carlo simulation based on the estimated rate matrices. The inset shows two exemplary force extension curves obtained from SMD (red) and kinetic Monte Carlo simulations (green). (e)~Mean rupture force $\bar{F}$ as a function of the loading rate $\mu=\kappa v$. The shaded areas indicate the $2\sigma$-confidence intervals.}
\label{fig:mean}
\end{figure*}

As mentioned above, SMD simulations do not allow to use pulling speeds as small as those used experimentally. In order to be able to determine transition times $\tau^\ast$ for such small values of $v$, we construct a series of 41 MSMs from MD simulations biased by a \emph{constant} potential $U^{(k)}=\frac{\kappa}{2}(\ree-\rpl^{(k)})^2$ characterized by the separation $\rpl^{(k)}$ with $k=1,\dots,41$. The configuration space of the catenane dimer is discretized into a finite set of states, $\ree^i$ with $i=1,\dots,60$. The dynamics between these states is governed by the transition rates $W^{(k)}_{ij}$ (from state $i$ to state $j$) depending on the biasing potential. The static free energy landscape along the order parameter $\ree$ obtained from the MSMs is depicted in Figure~\ref{fig:mean}(a) for selected values of $\rpl^{(k)}$. Note that all curves are shifted such that at $\ree=1.48\unit{nm}$ the respective free energy is zero. For $\rpl=2.8\unit{nm}$ the global minimum coincides with the closed state ($\ree<1.5\unit{nm}$), while the remaining part of the free energy is rather flat ($\ree>1.7~$nm). For $\rpl\geq3.1\unit{nm}$, on the other hand, the global minimum is shifted toward the open state ($\ree>1.8\unit{nm}$) with the closed state still exhibiting a distinct local minimum. Intermediate values of $\rpl$ illustrate the transition between both cases.

We now turn to extracting dynamic quantities from the MSMs. First, we consider the time-dependent average end-to-end distance, 
\begin{equation}
  \mean{\ree^\text{MSM}(t)} = \sum_i \ree^ip_i(t),
\end{equation}
with the sum over all discrete states, where $\ree^i$ is the end-to-end distance calculated for the $i$th state, and $p_i(t)$ is its probability at time $t$. To calculate this probability, we approximate the transition rates $W_{ij}(t)\approx W^{(k)}_{ij}$ through a step-wise profile with $\rpl^{(k)}\leqslant\rpl(t)<\rpl^{k+1}$ (for an illustration see Figure~\ref{fig:msm}). From an equilibrated initial state at $t=0$, the probability $p_i(t)$ is then obtained through Eq.~(\ref{eq:master_approx}) with intermediate times $t_k=(\rpl^{(k)}-\rpl^{(0)})/v$ and final time $t_K=t$. In Figure~\ref{fig:mean}(b) we show the resulting average end-to-end distance for $v=0.01~$m/s, comparing it directly to the average of 22 SMD trajectories conducted at the same speed. Both curves show excellent agreement, demonstrating that $\mean{\ree^\text{MSM}(t)}=\mean{\ree ^\text{SMD}(t)}$. Therefore, the dynamics in configuration space is perfectly described by a Markovian stochastic process. We note that a similar agreement is obtained for pulling speeds up to $v=0.1\unit{m/s}$.

We fit the average end-to-end distance curves with the function
\begin{equation}
  \label{eq:fit}
  \mean{\ree^\text{fit}(t)} = 
  \frac{a}{2}\left\{\tanh\left[(t-\tau^\ast)/\Delta\tau\right]+1\right\} + bt + c
\end{equation}
with four fit parameters: $a$ is the distance between open and closed state, $b$ is the slope away from the transition, and $c$ is an (irrelevant) offset. Moreover, $\tau^\ast$ yields the average unfolding time and $\Delta\tau$ is a measure for the temporal spreading of the stochastic unfolding events. In Figure~\ref{fig:mean}(c), we plot the unfolding rate $1/\tau^\ast$ as a function of pulling speed $v$. The unfolding rates for $v =\{0.01,0.1\}~$m/s following the MSM approach are in perfect agreement with the unfolding rates obtained from SMD simulations. Note that the fitting error is smaller than the symbols.
It is evident from Figure~\ref{fig:mean}(c) that the transition rates obtained from the MSM and the SMD techniques show an excellent overlap for one decade of pulling speeds. For the chosen parameters, this overlap is in the range of $0.01\unit{m/s}\le v\le 0.1\unit{m/s}$. Smaller pulling velocities would require unaffordable long simulation times in case of the SMD simulations and pulling speeds larger than $0.1~$m/s, on the other side, become problematic for the Markov state modeling approach, for the limitations of which we provide a detailed discussion in Sec.~\ref{sec:disc}. Varying the value for the force constant $\kappa$ the range of overlap of the two methods could be broadened~\cite{G86}.

The usual analysis of the results of SMD simulations and of force spectroscopy data consists in the determination of the rupture forces from force versus extension (FE) curves. We calculated such FE curves from individual $\ree(t)$ versus $t$ curves obtained from the MSMs via standard kinetic Monte Carlo (KMC) simulations.  The results are presented in Figure~\ref{fig:mean}(d) and the derived rupture force distributions agree excellently with those obtained directly from SMD simulations. The resulting force spectrum, the mean rupture force as a function of loading rate, is shown in Figure~\ref{fig:mean}(e).

\section{Discussion}
\label{sec:disc}

The rates $1/\tau^\ast$ are mean rates for the transition from the closed to the open configuration. 
Of course, for many systems (probably even containing polymeric linkers) the rates cannot be extracted straightfowardly from the end-to-end distance as in our model system.
Therefore, the standard procedure to compute transition rates is different.
One determines the rupture force from a single FE curve and repeats this procedure many times. 
This way, the distribution of rupture forces can be determined and from these one can then compute the transition rates~\cite{Dudko:2015}. The relation between the stochastic unfolding time $\tau$ and the rupture force $F$ derived from the pulling potential is given by
$F=\kappa[v\tau -(\ree(\tau)-\rpl^{(0)})]$, and thus one can either analyze the forces or the times. The mean force can also be determined from the dynamic strength, \emph{i.e.}, the mean FE curve~\cite{G67}, and we have the relation 
$\mean{F}=\kappa[v\tau^\ast -(\mean{\ree(\tau^\ast)}-\rpl^{(0)})]$. Using the (phenomenological) Bell model one has $\mean{F}\sim\ln{v}$ and therefore 
$1/\tau^\ast\sim\exp{(\mean{F}R^\dagger/k_\text{B}T)}\sim v$
with barrier position $R^\dagger$~\cite{Evans:2001,Dudko:2008}. We observe $1/\tau^\ast\sim v^\alpha$ with exponent $\alpha\simeq0.945$ [Figure~\ref{fig:mean}(c)], the slight deviation from the linear dependence might indicate deviations from the Bell model for fast pulling. 

In our approach, we can either compute average quantities like $\mean{\ree^\text{MSM}(t)}$ or distributions of rates or forces via KMC simulations of the MSMs. 
As an important example, we present the distribution of rupture forces, $p(F)$, in Figure~\ref{fig:mean}(d) as obtained from the individual FE curves.
We find excellent agreement between the distributions determined from SMD data and from the KMC simulations, a fact that substantiates the applicability of our coarse graining procedure.
The force spectrum, \emph{i.e.}, the mean force $\bar{F}=\int dFFp(F)$ versus loading rate $\mu=\kappa v$, is presented in Figure~\ref{fig:mean}(e). 
The slight curvature in the force spectrum presented in Figure~\ref{fig:mean}(e) indicates deviations from the Bell model for smaller loading rates. These deviations can be traced back to the reversible rebinding taking place in our model system~\cite{Friddle:2012, G80}.
The fact that the mean force tends to a constant value for very small loading rates indicates that the system resides in equilibrium.
This behavior is in accord with the experimental observations for small loading rates~\cite{G69}.

We finally discuss the limitations of the MSM approach to model the mechanical unfolding of small biomolecules. Theoretically, the pulling dynamics can be reconstructed for all pulling speeds. In practice, however, for certain pulling speeds the discretization is not fine enough, or the Markov assumption is not guaranteed to hold.

The upper limit for the pulling speed is determined by two factors. First, for small and large values of $\rpl^{(k)}$ the configuration space sampled by the MD simulations does not cover the full discretized state space anymore, \emph{i.e.}, transition rates linking states with high and low values of $R_\text{ee}$ are missing in the corresponding matrix $\mat W^{(k)}$ since these transitions are highly unlikely. However, for moderate and slow pulling speeds this is negligible since the system has time to traverse through states that have been sampled at intermediate values of $\rpl$.
When pulling too fast, on the other hand, the pulling time is much quicker than the implied time scales of the rate matrices. This causes, due to the missing transitions, the trapping of probability in states with low end-to-end distances at large $\rpl^{(k)}$. In principle, this limitation can be overcome by improved sampling. Second, the set of MSM rate matrices is estimated for a certain lag time. If the pulling speed is faster than a limit set by the inverse lag time, the Markov assumption is not guaranteed to hold anymore.

At small pulling speeds reversible binding and rebinding occurs, with the unfolding rate determined by barrier crossings (activated or Kramers' regime). In this regime the pulling only affects the slowest timescales of the current rate matrix with the other timescales corresponding to equilibrated intrawell dynamics. The (slight) qualitative change in the dependence on the pulling speed $v$ indicates that this regime is reached for $v\lesssim4\times10^{-6}\unit{m/s}$ with rate $1/\tau^\ast\sim v$ [left inset Figure~\ref{fig:mean}(c)]. Again, rare thermal transitions that would lead to a spontaneous opening of the dimer might not be sampled sufficiently in the MD simulations. Consequently, the rates estimated from the MSM approach in this regime are likely to underestimate the true unfolding rate.

We have applied our new technique of dynamic coarse graining to a well characterized model system that furthermore shows two-state behavior to an excellent approximation. We find a perfect match between our MSM based approach and the atomistic simulations. While for the \cal dimer the end-to-end distance is a sufficient order parameter to characterize the transition, more complex situations --including multi-dimensional order parameters which typically occur in biomolecular folding--  can be handled easily employing the well-developed tools for constructing Markov state models~\cite{MSMbook}.
Our strategy paves the way to investigate a number of interesting questions as it allows to cover a very broad range of pulling speeds.
The most obvious achievement lies in its broad applicability and it is now possible to directly compare the results of experiments recorded at vastly different pulling speeds. A very important issue is concerned with the possible speed-dependent change in the pathway of mechanical unfolding, in particular the crossover from thermal to mechanical 
unfolding~\cite{Stirnemann:2014,Zhuravlev:2016}.
Also the impact of dynamic disorder in the mechanical unfolding can be investigated over a huge range of rates and thus might allow to probe different regimes of this fascinating topic~\cite{Hyeon:2014}.
Furthermore, the crossover between the different regimes defined by the stiffness of the pulling device, 
$\kappa$, can be studied in detail.
We conclude with mentioning that our method opens many opportunities to directly compare atomistic simulations with force spectroscopy studies on experimentally relevant time scales.


\section*{Acknowledgements}

Financial support by the DFG via the collaborative research center TRR 146 (projects A7 and B3) is gratefully acknowledged. G.D. and K.S. acknowledge support by the DFG via Grant No. DI693/3-1.


\end{document}